\documentstyle[twoside,fleqn,espcrc2,epsf]{article}

\newcommand{\AmS}{{\protect\the\textfont2
  A\kern-.1667em\lower.5ex\hbox{M}\kern-.125emS}}

\title{Toward Precision Measurement of $B_K$ with
       Quenched Kogut-Susskind Quarks\thanks{presented by S. Aoki}}

\author{JLQCD Collaboration\\[2mm]
	S. Aoki\address{Institute of Physics, University of Tsukuba,
        Tsukuba, Ibaraki 305, Japan},
        M. Fukugita\address{Yukawa Institute for Theoretical Physics,
        Kyoto University, Kyoto 606, Japan},
        S. Hashimoto\address{National Laboratory for High Energy Physics (KEK),
        Tsukuba, Ibaraki 305, Japan},
	Y. Iwasaki$^{\rm a,d}$,
	K. Kanaya$^{\rm a,}$\address{Center for Computational Physics,
        University of Tsukuba, Tsukuba, Ibaraki 305, Japan},
	Y. Kuramashi$^{\rm c}$,
        H. Mino\address{Faculty of Engineering, Yamanashi University,
        Kofu 400, Japan},\\
	M. Okawa$^{\rm c}$,
	A. Ukawa$^{\rm a}$,
	T. Yoshi\'e$^{\rm a,d}$
}

\begin{document}

\begin{abstract}
We present a status report of our ongoing effort toward a precision
determination
of the kaon $B$ parameter with the Kogut-Susskind quark action
in quenched QCD.
Results for $B_K$ so far accumulated at $\beta =5.85$, 5.93, 6.0 and 6.2
corresponding to $a^{-1}=1.3-2.6$ GeV do
not exhibit the theoretically expected $O(a^2)$ behavior,
but are apparently more consistent with an $O(a)$ behavior.
\end{abstract}

\maketitle

\section{Introduction}

The kaon $B$ parameter is one of the crucial hadron weak matrix
elements for constraining the $CP$ violation phase of the
Cabibbo-Kobayashi-Maskawa matrix from experiments.  A pioneering
study was carried out by Gupta, Kilcup, Patel and Sharpe
(GKPS)\cite{STAG,sharpe0,sharpe1}, which laid a foundation for a lattice QCD
determination of this quantity with the Kogut-Susskind quark action.
A number of problems, however, are posed to warrant viability of the
lattice result in the continuum.
While some aspects have been examined previously
\cite{IFMOSU,lee}, further studies are needed
to establish the continuum value of $B_K$.
The statistics of GKPS are only 15 ($\beta$=6.0) to 24 ($\beta$=6.4)
configurations: as a consequence the error of $B_K$ at
$\beta=6.4$ is so large that an $O(a^2)$ dependence of $B_K$ on the
lattice spacing inferred from a theoretical analysis\cite{sharpe1} can not
be ascertained. Furthermore the lattice size of
$32^3\times 48$ employed at $\beta=6.2$ and 6.4
may be too small in the temporal direction.
We have therefore decided to carry out a new set of simulations
with improved statistics, incorporating advance in methods of
analyses.  Here we present a status report of this program which is
being carried out on VPP500/80 newly installed at KEK.

\section{Simulation method and run parameters}

We basically employ the formalism developed by GKPS
for calculating the $B$ parameter.
Following improvements are included in our simulations, however.

On each configuration we calculate $2^3=8$ quark propagators corresponding
to wall sources with
a unit source placed at one of the corners of each spatial cube.
This allows us to construct a wall source which creates only the
pseudoscalar meson in the Nambu-Goldstone channel\cite{8wall}.
In contrast the even-odd source employed in
previous studies\cite{STAG,sharpe0,sharpe1,IFMOSU} also creates vector mesons
that contaminate signals.  The sources are placed at the two edges
of the lattice, and the Dirichlet boundary condition is employed in the
temporal direction.  Gauge configurations are fixed to the Landau gauge.

Four-quark operators written in terms of Kogut-Susskind quark fields are
extended over a $2^4$ hypercube.  We employ
both gauge-invariant and non-invariant operators\cite{IFMOSU},
which differ by an
insertion of appropriate gauge link factors for the former.
The one-loop result for the lattice renormalization factors\cite{IS,SP}
is used to convert lattice values of $B_K$ to those in the continuum,
renormalized with the naive dimensional regularization(NDR) scheme.

A source of uncertainty in the conversion is the
choice of the gauge coupling constant used for the renormalization
factors.  We take the mean-field improved $\overline{MS}$ coupling
constant at the cut-off scale $\mu=\pi/a$\cite{elkahdra}.
Correspondingly we improve
the operators\cite{lepagemackenzie} through the replacement
$\chi\rightarrow\sqrt{u_0}\chi$ and
$U_\mu \rightarrow u_0^{-1} U_\mu$, where $\chi$ is the Kogut-Susskind quark
field, $U_\mu$ the link variable and $u_0 =P^{1/4}$ with $P$ the average
plaquette.  Modifications of renormalization factors due to this
replacement\cite{IS,SP} are also taken into account.

We note that we have ignored the mixing of non-Nambu-Goldstone
four-quark operators in the present analysis.  Studies of this mixing
effect, which is believed to vanish in the continuum limit,
are left for future work.

\begin{table}[t]
\setlength{\tabcolsep}{0.2pc}
\caption{Run parameters}
\label{tab:run}
\begin{tabular}{lllll}
\hline
$\beta$ & 5.85 & 5.93 & 6.0 & 6.2 \\
\hline
\\[-3mm]
$L^3T$ & $16^332$ & $20^340$ & $24^340$ &$32^364$ \\
\#conf. & 60 & 50 & 50 & 40 \\
$a^{-1}$ & 1.34(3) & 1.58(4) & 1.88(5)  & 2.62(9) \\
$m_qa$ & 0.01-0.04 & 0.01-0.04 & 0.01-0.03 & 0.005-0.02 \\
$m_sa/2$ & 0.0197 & 0.0156 & 0.0125 & 0.0089\\
fit & 10-20 & 12-26 & 14-24 & 20-42 \\
\hline
\end{tabular}
\vspace*{-7mm}
\end{table}

Parameters of runs we have so far carried out are listed in
Table~\ref{tab:run}. Gauge configurations are generated with the 5-hit
heatbath algorithm at 2000($\beta\leq 6.0$) or 5000($\beta=6.2$) sweep
intervals.
The physical scale of lattice spacing is determined from the $\rho$ meson
mass in the $VT$ channel.
Values of $\beta$ are chosen so that the lattice spacing is roughly
equally spaced, and the lattice size is increased for weaker coupling to
keep the physical lattice size approximately constant at
$La\approx 2.4-2.5$ fm.
A similar consideration is made for the choice of the fitting interval
in extracting $B_K$ from kaon Green's functions.
Quark mass is taken in steps of $0.01$ for $\beta\leq 6.0$ and 0.005 at
$\beta=6.2$.   While we have evaluated $B_K$ for non-degenerate strange
and down quark masses,  we shall restrict ourselves to the degenerate
case in the present analysis.
The strange quark mass yielding the observed ratio $m_K/m_\rho = 0.648$
for this case is also given in Table~\ref{tab:run}.
We estimate errors by a single elimination jackknife procedure.

\section{Comparison with previous work}

\begin{figure}
\centerline{\epsfxsize=7.5cm \epsfbox{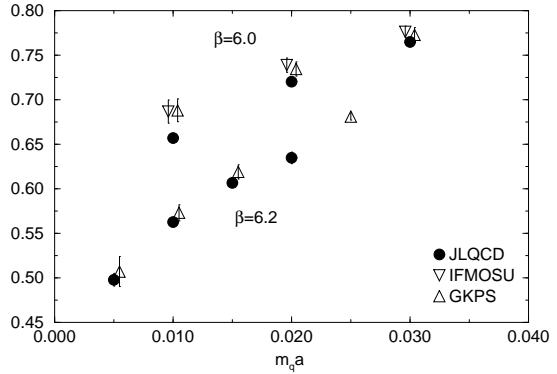}}
\vspace*{-12mm}
\caption{Comparison of lattice value of $B_K$ for gauge non-invariant
operator at $\beta = 6.0$ and $6.2$. Data for $\beta=6.2$ are shifted downward
by 0.05 to avoid overlap.}
\label{comp60a}
\vspace*{-5mm}
\end{figure}

In Fig.~\ref{comp60a} lattice values of $B_K$ for the gauge
non-invariant operator at $\beta =6.0$ and $6.2$ are plotted.
At $\beta =6.2$ our results are consistent with those of
GKPS\cite{sharpe0,sharpe1} at the level of one standard deviation.
On the other hand, our values at $\beta=6.0$ are systematically lower than
those of GKPS and Ishizuka {\it et al.}(IFMOSU)\cite{IFMOSU}
obtained on the same size of lattice.

\begin{figure}[t]
\centerline{\epsfxsize=7.5cm \epsfbox{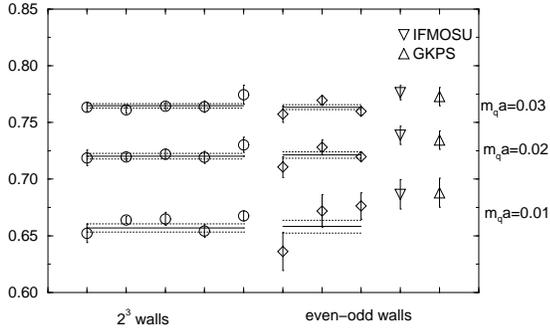}}
\vspace*{-12mm}
\caption{Lattice value of $B_K$ at $\beta=6.0$ calculated for
10 configurations with $2^3$ (circles) and even-odd (diamonds) source as
compared with previous results obtained with even-odd source
(triangles).  Solid lines are
values for the full ensemble with errors indicated by dotted lines. }
\label{comp60b}
\vspace*{-5mm}
\end{figure}

In order to understand the discrepancy, we note that GKPS and IFMOSU
respectively used 15 and 10 configurations as compared to our 50,
and that both groups employed the even-odd wall source, while
we used the $2^3$ source.
For a comparison under the same condition, we divided our
ensemble into 5 sets of 10 configurations and recalculated $B_K$ for
each set. In addition we made a new calculation for
3 sets of 10 configurations using the even-odd wall source.
The eight values of $B_K$ thus obtained are plotted in
Fig.~\ref{comp60b} and
are compared with those of IFMOSU and GKPS.
Solid lines represent values obtained for the full ensemble for each type
of source.

We note that, while results for the full ensemble are
consistent between the two types of sources, both errors and fluctuations of
central values are significantly reduced
with the $2^3$ source, demonstrating the advantage of this method.
Since values of GKPS and IFMOSU are consistent with ours for the even-odd
source within one to two
standard deviations, we conclude that previous results have overestimated
the value of $B_K$ at $\beta=6.0$ due to insufficient statistics.

\section{Quark mass dependence}

\begin{figure}
\centerline{\epsfxsize=7.5cm \epsfbox{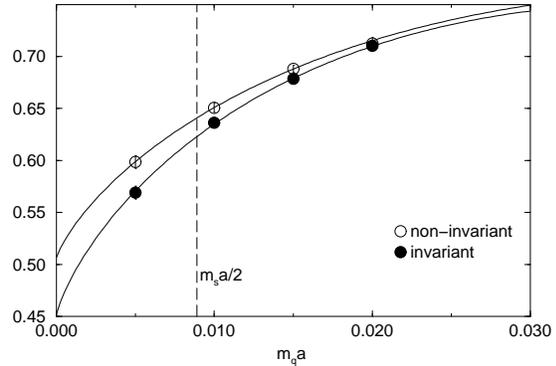}}
\vspace*{-12mm}
\caption{Mean-field improved $B_K$(NDR, 2GeV) as a function of $m_qa$
at $\beta = 6.2$ on a $32^3\times64$ lattice.
Solid lines are fits to (\protect\ref{eq:chiral}) and dashed line corresponds
to physical value of $m_K/m_\rho$.}
\label{BKmq}
\vspace*{-5mm}
\end{figure}

Extraction of the physical value of $B_K$ requires an interpolation of
results to the quark mass corresponding to the physical kaon.
We find a suggestive result that the quark mass dependence of our data is
well described by the form predicted by chiral perturbation theory, which
we employ for the above purpose.
For the case of degenerate strange and down quark treated in the
present analysis, the formula is given by\cite{sharpe2}
\begin{equation}
B_K = B\left[ 1- 3 c (m_qa)\log (m_qa) + b (m_qa) \right],
\label{eq:chiral}
\end{equation}
where $c=\alpha/(4\pi f_\pi)^2$ with $\alpha=(m_Ka)^2/m_qa$ and $b$ is an
unknown constant.  A representative example of the fit taking $B, c$ and $b$
as parameters is plotted in Fig.~\ref{BKmq} together with the quark mass
corresponding to $m_K/m_\rho=0.648$ (dashed line).
Several uncertainties exist, however, to take the agreement as evidence
for the presence of chiral logarithm in our data.  The pion decay constant
$f_\pi$ estimated from the fitted values of $c$ and the spectrum results
for the slope $\alpha$ turned out to have a large value of $150-300$ MeV with
$10-20$\% error as compared to the experimental value of 93 MeV.
Also a quadratic form without logarithm fits our data equally well.

\section{Results for $B_K$}

\begin{figure}[t]
\centerline{\epsfxsize=7.5cm \epsfbox{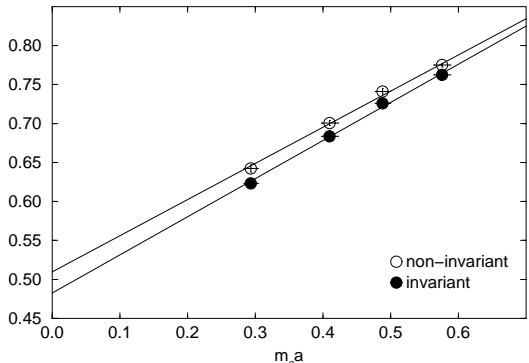}}
\vspace*{-12mm}
\caption{Mean-field improved $B_K$({\rm NDR}, 2GeV) as a function of
$m_\rho a$, together with linear fits (solid lines).}
\label{BKmrho}
\vspace*{-5mm}
\end{figure}

We plot our results for $B_K$ renormalized at $\mu=2$ GeV in the NDR scheme
as a function of $m_\rho a$ in Fig.~\ref{BKmrho} both for gauge invariant
(filled circles) and non-invariant (open circles) operators.
We observe that the data do not exhibit the theoretically predicted $O(a^2)$
dependence\cite{sharpe1}, and are apparently more consistent with a linear
behavior,
at least over the range $m_\rho a=0.57-0.29$ (or $1/a=1.3-2.6$ GeV from
$m_\rho$) corresponding to $\beta=5.85-6.2$.  If we assume a linear form,
we obtain $B_K(a)=0.4825(83)+0.489(17) m_\rho a$ and
$0.5097(83)+0.464(17) m_\rho a$
for the gauge invariant and non-invariant operators.

We have varied the analysis procedure to see to what
extent the values of $B_K$ are modified:
(i) Instead of $m_K/m_\rho$ we
examined the choice $m_\phi/m_\rho$ for estimating the strange quark mass.
This only resulted in a common downward shift of the four values of $B_K$ by
about $0.02$, and hence does not change a linear dependence on the
lattice spacing.  (ii) We also replaced the mean-field improved
$\overline{MS}$ coupling constant $g_{\overline{MS}}(\pi/a)$ in the
renormalization factors by the
running coupling constant at $\mu=2$ GeV evaluated by the two-loop
renormalization group evolution starting from $g_{\overline{MS}}(\pi/a)$.
This increases the value of $B_K$ by about 0.03-0.05 without, however,
changing the linear behavior.

Let us comment on the fact that our results for $B_K$
obtained with the two types of operators do not agree with
each other and that the discrepancy does not become smaller toward
the continuum limit.
Possible sources of the discrepancy are an operator dependence of finite
lattice spacing effects, which disappears in the continuum limit, and
$O(g^4)$ corrections in the
renormalization factors which are not included in our analysis.
Naively one would expect the magnitude of
$O(g^4)$ corrections to also diminish toward the continuum limit.
This is not the case, however, for $B_K$ renormalized at a fixed scale
$\mu=2$ GeV, since the gauge coupling constant $g(\mu)$ governing these
corrections is independent of the lattice spacing.
We have checked that
increasing the scale $\mu$ leads to a smaller discrepancy.
Thus the difference of $B_K$ for the two types of operators
extrapolated to the continuum limit provides
an estimate of the magnitude of $O(g^4)$ corrections.

Finally we remark that an $O(a^2)$ dependence is not completely ruled out
from our data if there exists an appropriate $O(a^3)$ or $O(a^4)$ term
which just brings the curve to behave as if it is linear over a range of
lattice spacing including that explored in our simulation.
If this rather exceptional case happens, the value of $B_K$ extrapolated to
the continuum limit $a=0$ would increase by about 0.05.

\section{Summary}

In this report we have described our effort toward improving
previous work on the kaon $B$ parameter with the Kogut-susskind quark action
in quenched QCD.  The results obtained at four values of $\beta$
show that $B_K$ falls linearly with the lattice spacing over the
range $1/a=1.3-2.6$ GeV. If the theoretical prediction of an $O(a^2)$
dependence is
valid, this means that such a behavior will become apparent only for
smaller values of lattice spacing.  We plan to carry out
a simulation at $\beta=6.4-6.5$ employing the lattice size of $40^3\times 96$.

\end{document}